\documentclass[a4paper,aps,pra,showpacs,superscriptaddress,twocolumn,nofootinbib]{revtex4-1}      
\usepackage[utf8]{inputenc}  
\usepackage[T1]{fontenc}     
\usepackage[british]{babel}  
\usepackage[sc,osf]{mathpazo}
\usepackage[scaled=0.86]{berasans}  
\usepackage[scaled=1.03]{inconsolata} 
\usepackage[colorlinks,linkcolor=magenta,urlcolor=blue]{hyperref}  
\usepackage{graphicx} 
\usepackage[babel]{microtype}  
\usepackage{amsmath,amssymb,amsthm,bm,mathtools,amsfonts,mathrsfs,bbm} 
\usepackage{xspace}  
\usepackage{centernot}
\usepackage{pgfplots}  

\DeclareMathOperator{\Tr}{Tr}

\newcommand{\bra}[1]{\left\langle #1 \right|}
\newcommand{\ket}[1]{\left| #1 \right\rangle}

\newcommand{\ketbra}[2]{\left|#1\middle\rangle\middle\langle#2\right|}

\newcommand{\ba}{\begin{eqnarray}}
\newcommand{\ea}{\end{eqnarray}}

\newcommand{\figref}[1]{Fig.~\ref{#1}}

\def\be{\begin{equation}}
\def\ee{\end{equation}}
\mathtoolsset{centercolon}

\begin{document}
\title{Small quantum absorption refrigerator in the transient regime: \\ time scales, enhanced cooling and entanglement}
\author{Jonatan Bohr Brask}
\affiliation{Département de Physique Théorique, Université de Genève, 1211 Genève, Switzerland}
\author{Nicolas Brunner}
\affiliation{Département de Physique Théorique, Université de Genève, 1211 Genève, Switzerland}

\date{\today}  

\begin{abstract}
A small quantum absorption refrigerator, consisting of three qubits, is discussed in the transient regime. We discuss time scales for coherent dynamics, damping, and approach to the steady state, and we study cooling and entanglement. We observe that cooling can be enhanced in the transient regime, in the sense that lower temperatures can be achieved compared to the steady-state regime. This is a consequence of coherent dynamics, but can occur even when this dynamics is strongly damped by the dissipative thermal environment, and we note that precise control over couplings or timing is not needed to achieve enhanced cooling. We also show that the amount of entanglement present in the refrigerator can be much larger in the transient regime compared to the steady-state. These results are of relevance to future implementations of quantum thermal machines.
\end{abstract}

\maketitle

Recently, the study of small self-contained quantum thermal machines has received growing interest; see \cite{review1,review2,Goold15,anders15} for recent reviews. Such machines typically consist of only a few quantum levels, hence can be considered 'small' quantum systems (in terms of Hilbert space dimension). Moreover, these machines are termed self-contained (or autonomous) as they function without any source of work or external control, but use only heat baths at different temperatures. The simplicity of these models makes them an ideal testbed for investigating quantum thermodynamics \cite{Goold15}.

First works in this area go back to the study of the thermodynamics features of lasers \cite{scovil59}. Since then, many designs have been proposed and studied (see e.g. \cite{GevKos96,PalKosGor01,luis,ralph}), among these a quantum absorption refrigerator consisting of three qubits \cite{linden10}. The efficiency of this machine \cite{skrzypczyk11} and more general performance bounds \cite{correa13,woods15} were discussed. The basic functioning and fundamental limits of the fridge can be captured via the concept of virtual qubits \cite{brunner12}. Moreover, quantum entanglement was shown to appear in this model, and to enhance cooling in certain regimes \cite{brunner14}. Possibilities for experimental implementations \cite{chen12,venturelli13,bellomo} were also discussed.

So far, most works have discussed quantum absorption refrigerators in the steady-state regime, giving a detailed characterization of its physical properties. On the other hand, the transient regime remains basically unexplored so far. The latter is however of interest. First, from a conceptual point of view, it is relevant to understand the approach to equilibrium. Second, from a more applied point of view, it is natural to ask how fast cooling can be achieved, and what the timescale for reaching equilibrium is. The study of quantum effects, such as entanglement and coherence in the transient regime is also an interesting issue.

Here we investigate the physics of a quantum absorption refrigerator in the transient regime. We focus on the three-qubit quantum fridge model of Ref. \cite{linden10}, characterizing time scales, cooling properties and entanglement. First, the time scales for coherent dynamics, damping, and decay to the steady state in terms of the bath coupling and interaction strengths are discussed. Then we observe that cooling can be enhanced in the transient regime via the coherent dynamics of the system. Specifically, it is possible to bring the object to be cooled to a temperature which is much lower than its steady-state temperature. As discussed recently in Ref. \cite{mitchison15}, this is a genuinely quantum feature, which highlights the advantage offered by quantum refrigerators over purely classical ones. Moreover, we find that neither precise timing nor accurate control of the coupling strengths are necessary for taking advantage of this cooling enhancement. Finally, we also observe that the amount of entanglement that can be achieved in the model can be much larger in the transient regime compared to the steady-state regime, which is again a consequence of the coherent dynamics. We believe that the present results opens novel questions for quantum thermal machines and may be of relevance to future possible practical implementation of these ideas.

\section{Model and master equation}

\begin{figure}[t]
\includegraphics[width=0.9\columnwidth]{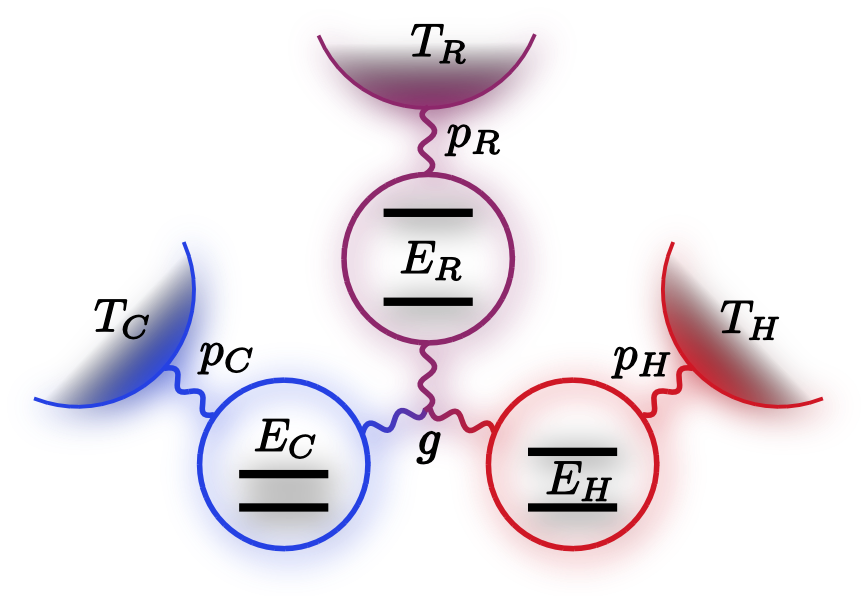}
\caption{(Color online) Three-qubit quantum thermal machine. Three two-level quantum systems (qubits) with energy gaps $E_C$, $E_R$, and $E_H$ are coupled to separate thermal reservoirs at temperatures $T_C$, $T_R$, and $T_H$ with coupling rates $p_C$, $p_R$, and $p_H$. The qubits interact collectively as described in the text, with interaction strength $g$. When operating as a fridge, the machines cools qubit $C$, i.e. brings it to a temperature below $T_C$.}
\label{fig.threeqfridge}
\end{figure}

We consider the model of a three-qubit quantum absorption refrigerator discussed in Refs \cite{linden10,skrzypczyk11} and sketched in Fig. 1. The ground and excited state of qubit $i$ are denoted by $\ket{0}_i$ and $\ket{1}_i$ and the energy gap $E_i$. The free Hamiltonian for the system is thus given by
\begin{equation}
H_0 =  \sum_{i\in\{C,R,H\}} E_i \ket{1}_i \bra{1}  .
\end{equation}
We set $E_R=E_C + E_H$ and $E_C \neq E_H$, and thus the states $\ket{010}$ and $\ket{101}$ are degenerate in energy (we use the ordering $CRH$). Moreover we consider an interaction Hamiltonian
\begin{equation}
\label{eq.Hint}
H_{int} = g ( \ketbra{010}{101} + \ketbra{101}{010} ) ,
\end{equation}
where $g$ is the coupling strength. We focus on the weak coupling regime, $g \ll E_i$, hence all transitions (apart from $\ket{010} \leftrightarrow \ket{101}$) are exponentially suppressed. 

Each qubit is weakly coupled to a thermal bath. The first qubit is connected to the coldest bath, at temperature $T_C$. The third qubit is connected to the hottest bath, at temperature $T_H$. The middle qubit is connected to a bath at intermediate temperature $T_R$, with $T_C \leq T_R \leq T_H$. The interaction between each qubit and its bath is modeled by a simple reset model (see e.g. \cite{skrzypczyk11}) where thermalisation happens through rare but strong events. At every time step, each qubit $i$ is either reset to a thermal state $\tau_i$ at the temperature of its bath with a small probability, or left unchanged. The evolution of the qubit states is thus given by the master equation
\begin{equation}
\label{eq.master}
\frac{\partial\rho}{\partial t} = - i [H_0 + H_{int},\rho] + \sum_{i\in\{C,R,H\}} p_i (\Phi_i(\rho) - \rho)
\end{equation}
where $p_i$ is the thermalisation rate for qubit $i$ and
\begin{equation}
\Phi_i(\rho) = \tau_i\otimes\Tr_i(\rho)  
\end{equation}
where $\Tr_i$ denotes the partial trace over qubit $i$ and the tensor product is to be taken at position $i$. The thermal states are given by $\tau_i = r_i \ketbra{0}{0} + (1-r_i)\ketbra{1}{1}$ with 
\begin{equation}
\label{eq.thermalpop}
r_i = \frac{1}{e^{-E/T_i} + 1}
\end{equation}
where $T_i$ is the reservoir temperature for qubit $i$ (throughout the paper we set $k_B = 1$). We note that the master equation applies in the perturbative regime where $p_i,g \ll E_C,E_H$ and $p_i \ll 1$ (see e.g.\cite{skrzypczyk11}). In this case, thermalization events between a given qubit and the heat bath associated with the other two qubits are second order events which can be safely neglected.

In \cite{linden10} it was shown in detail how the machine can operate as a refrigerator, considering the steady-state regime. Briefly, when the reduced states of each qubit are diagonal, we can associate temperatures $T_c$, $T_r$, $T_h$ with them via \eqref{eq.thermalpop}. This is the case in the steady state, where the cooling effect can be simply understood in the virtual qubit picture developed in \cite{brunner12}. The interaction $H_{int}$ effectively puts the cold qubit in thermal contact with a virtual qubit spanned by the levels $\ket{01}_{RH}$ and $\ket{10}_{RH}$ of the other two qubits. In the absence of interaction, the temperature of this virtual qubit is
\begin{equation}
\label{eq.Tv}
T_V = \frac{E_C}{E_R/T_R - E_H/T_H}.
\end{equation}
When $0 \leq T_V < T_C$, the machine acts as a refrigerator, in the sense that the cold qubit will be cooled below its bath temperature, i.e.~$T_V < T_c < T_C$.

\section{Solving the master equation}

Previous works have discussed the steady state solution of the above model in great detail, see e.g. \cite{linden10,skrzypczyk11,brunner14}. Here our focus is different as we are interested in the transient regime. 

We start by pointing out that \eqref{eq.master} is linear in $\rho$ and can thus be recast as a matrix differential equation 

\begin{equation}
\label{eq:differ}
\frac{\partial \mathbf{v}}{\partial t} = A \mathbf{v} + \mathbf{u}, 
\end{equation}
where $\mathbf{v}$ is simply a rewrapping of the density matrix $\rho$ to a vector. The matrix $A$ and vector $\mathbf{u}$ depend on the parameters $E_i$, $g$, $p_i$, $T_i$, and encode the right-hand side of \eqref{eq.master}. The steady-state solution is given by $\mathbf{v}_\infty = - A^{-1}\mathbf{u}$. At intermediate times, \eqref{eq:differ} is solved by $\mathbf{v}(t) = \mathbf{v}_h(t) + \mathbf{v}_\infty$, where $\mathbf{v}_h(t)$ is a general solution to the homogeneous equation with $\mathbf{u} = 0$, which can be obtained by diagonalising $A$. Denoting the eigenvalues and eigenvectors of $A$ by $\lambda_j$ and $\mathbf{e}_j$ respectively, one has
\begin{equation}
\label{eq.vht}
\mathbf{v}_h(t) = \sum_j c_j e^{\lambda_j t} \mathbf{e}_j ,
\end{equation}
where the coefficients $c_j$ are determined by the initial condition that $\mathbf{v}(0)$ matches the given input state. Note that the real part of all $\lambda_i$ must be negative such that $\mathbf{v}_h(t)$ vanishes at long times and the steady state $\mathbf{v}_\infty$ is recovered.

It turns out that diagonalizing $A$ in full generality is challenging. Nevertheless, for fixed values of the parameters $E_i$, $g$, $p_i$, $T_i$, we can easily obtain the time-dependent solution to \eqref{eq:differ} and hence the state $\rho(t)$ for any given initial condition $\rho(t=0)=\rho_0$. We also know the steady-state $\rho_\infty$ for given parameters (using the method above, or from \cite{skrzypczyk11}).

\section{Characterizing the transient regime}

We are now in position to start discussing the physical properties of the three-qubit fridge in the transient regime. We will first consider the time scales involved in the approach to the steady state, and then look at cooling and entanglement in the transient regime.

\begin{figure*}[t!]
\includegraphics[width=0.75\linewidth]{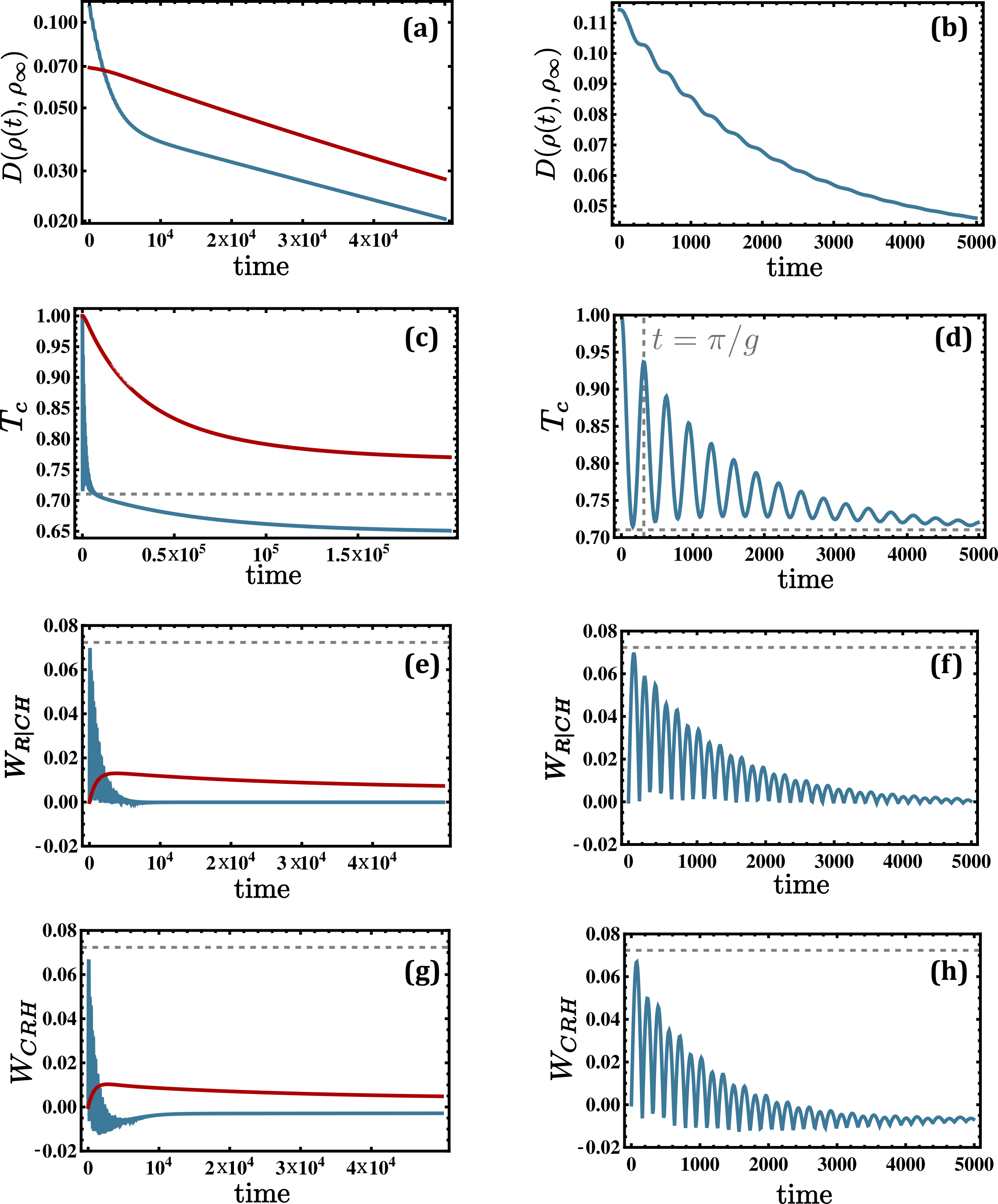}
\caption{(Color online) Plots vs.~time of \textbf{(a)-(b)} the distance to the steady state $D(\rho(t), \rho_\infty)$, \textbf{(c)-(d)} the cold qubit temperature $T_c(t)$, \textbf{(e)-(f)} bipartite entanglement $W_{R|CH}(t)$, and \textbf{(g)-(h)} genuine tripartite entanglement $W_{CRH}(t)$. We use two sets of parameters given by $E_C=1$, $E_H=100$, $T_C = T_R = 1$, $T_H = 100$, $p_C = p_H = 10^{-5}$, $p_R=10^{-3}$, and $g=10^{-2}$ (blue curves), $g=10^{-4}$ (red curves), and the system is initially in a thermal state with each qubit equilibrated to its bath. In (c) and (d) the minimal cold qubit temperature attainable by unitary dynamics alone is indicated (dashed horizontal), and in (d) also the period of the unitary dynamics (dashed vertical). In (e)-(h) the maximal entanglement extractable from the initial state by the unitary dynamics alone is indicated (dashed).}
\label{fig.transplots}
\end{figure*}

Throughout the following we will consider a fixed, natural initial state, namely a thermal state where each qubit is at equilibrium with its bath
\begin{equation}
\tau = \tau_C\otimes\tau_R\otimes\tau_H .
\end{equation}
This is the equilibrium state corresponding to the free Hamiltonian $H_0$ that one has before the interaction is turned on. Initially there is thus no entanglement, and each qubit is at the temperature of its respective bath. Since the state is diagonal in the basis of $H_0$ the only elements of the density affected by the unitary evolution, once the interaction is turned on, are those in the degenerate subspace affected by $H_{int}$. In particular, since the interaction with the baths do not generate coherences within each individual qubit, the only off-diagonal elements will be $\ket{010}\bra{101}$ and $\ket{101}\bra{010}$. This also implies that the reduced state of each qubit is diagonal at any time, and we can thus associate temperatures with them.

In addition to fixing the initial state, we will focus on the role played by the relative strengths of the couplings $p_i$ and the interaction $g$.  Without loss of generality, we can set the energy scale such that $E_C=1$. We also fix the other qubit energy and the bath temperatures. For simplicity we set $T_C = T_R$ (this also analogous to a natural situation where the cold bath is at room temperature). As long as the qubits are sufficiently far from each other we can model their baths as independent \cite{mitchison15,palma96}.

\subsection{Quantities of interest}

Before moving on, we define the quantities of interest we will use below to study the approach to the steady state, the cooling, and the entanglement in the transient regime.

To examine the approach to equilibrium, one needs a natural notion of distance from the steady state, and we take the trace distance
\ba
D(\rho(t), \rho_\infty)  = \frac{1}{2} ||  \rho(t)- \rho_\infty  ||_1 ,
\ea
which measures the distinguishability between the two states ($||\cdot||_1$ is the trace norm). Specifically, this quantity has an operational meaning, as it is equal to the classical trace distance between the probability distribution of the measurement outcomes for the optimal quantum measurement distinguishing between $\rho(t)$ and $\rho_\infty $ \cite{nielsenchuang}.

For cooling, we note that starting from the product thermal state $\tau$, the only off-diagonal elements of $\rho(t)$ at any time are those in the degenerate subspace. Therefore the reduced state of each qubit is diagonal, and using \eqref{eq.thermalpop} we can associate temperatures $T_c$, $T_r$, $T_h$  with them. When operating as a refrigerator the machine cools the cold qubit such that $T_c < T_C$.

Finally, for the entanglement in the three-qubit system, there are several bipartitions and quantities one may study. Entanglement in the steady state was discussed in Ref.~\cite{brunner14}, and all types of three-qubit entanglement were shown to occur in various regimes. Following this work, we evaluate entanglement along a given bipartition (e.g. qubit 1 versus qubits 2 and 3), or genuine tripartite entanglement, using a class of entanglement witnesses developed in \cite{guhne10,huber10} which allow one to fully characterize the entanglement of states of the form $\rho(t)$ for initial state $\tau$. Specifically we consider witnesses of the form
\ba
\label{eq.entwitness}
W_{\mathcal{S}}(\rho(t))=\label{witness} 2 \left( |\rho_{3,6}|- \sum_{j\in\mathcal{S}}\sqrt{\rho_{j,j} \rho_{9-j,9-j}} \right) \leq 0 ,
\ea
where $\rho_{i,j}$ denotes elements of the density matrix $\rho(t)$ and the set $\mathcal{S}$ depends on the partition and type of entanglement one is interested in. When inequality \eqref{witness} is violated, its left-hand side gives the concurrence \cite{wootters} of $C|RH$ ($\mathcal{S}=\{2\}$), $R|CH$ ($\mathcal{S}=\{1\}$), $CR|H$ ($\mathcal{S}=\{3\}$) or the genuine multipartite concurrence (see Refs.~\cite{ma11,wu12}) for $\mathcal{S}=\{1,2,3\}$. When inequality \eqref{witness} holds, no entanglement is present on the given bipartition, as the witness provides a necessary and sufficient condition for biseparability \cite{hashemi12}.

\subsection{Time scales}

In \figref{fig.transplots} we show representative plots illustrating the transient behaviour of the distance to the steady state, the entanglement, and cooling. We observe the following general behaviour. When the interaction strength $g$ exceeds the bath couplings $p_i$, all of the observed quantities initially oscillate at a frequency of approximately $g/\pi$ (see \figref{fig.transplots}). This is intuitive since this is the time scale of the system dynamics in the absence of dissipation (as can be seen from the interaction Hamiltonian). Hence for weak bath coupling, we expect to see such oscillations until dissipation becomes dominant. 

More specifically, from \eqref{eq.vht} the time scales are determined by the eigenvalues $\lambda_j$ of the matrix $A$. Since starting from the initial state $\tau$, $\rho(t)$ always features a single off-diagonal element, we can take $A$ to be $9 \times 9$. We observe the following general properties of the spectrum of $A$. Only two of the eigenvalues are complex (conjugates of each other) $\lambda_{cp}$, $\lambda_{cp}^*$. Moreover, all eigenvalues have negative real parts (ensuring convergence to the steady state). The largest (numerically smallest) eigenvalue $\lambda_{max}$ is always real. 

\begin{figure}[t!]
\includegraphics[width=0.8\linewidth]{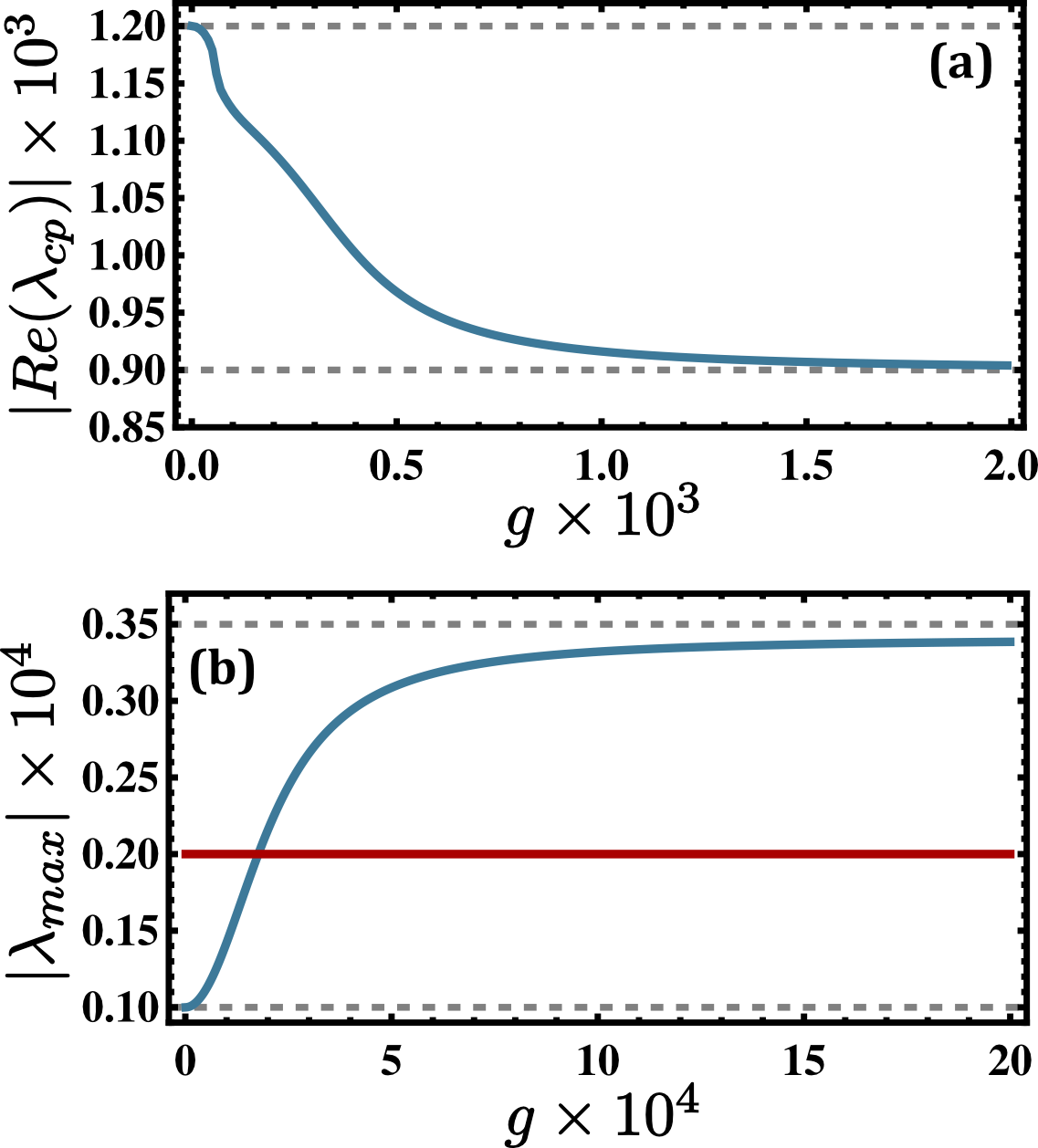}
\caption{(Color online) \textbf{(a)} Damping rate of coherent dynamics vs.~interaction strength for $E_C=1$, $E_H=100$, $T_C=1$, $T_R=1$, $T_H=100$, $p_C=10^{-4}$, $p_R=10^{-3}$, $p_H=10^{-4}$. The limiting values $p_C+p_R+p_C$ and $3(p_C+p_R+p_C)/4$ for small and large g respectively are indicated (dashed). \textbf{(b)} Asymptotic decay rate vs.~interaction strength for $E_C=1$, $E_H=100$, $T_C=1$, $T_R=1$, $T_H=100$, $p_C=10^{-4}$, $p_R=10^{-3}$, $p_H=10^{-5}$ (blue) and $p_C=p_H=2\times 10^{-5}$ (red). The limiting values $\min\{p_C,p_R,p_H\}=p_H$ and $\sim p_H+p_C/4$ for the case of all $p_i$ different are indicated (dashed).}
\label{fig.lambdag}
\end{figure}

The coherent dynamics is characterized by oscillations, the frequency of which is given by the imaginary part of $\lambda_{cp}$,  while the time scale for damping is given by the real part. When $g$ is small compared to the couplings $p_i$, the real part of $\lambda_{cp}$ is $p_C+p_R+p_H$, while for large $g$ it is $3(p_C+p_R+p_H)/4$. A typical example is shown in \figref{fig.lambdag}(a). The oscillations are therefore damped out in a time that scales as the inverse of $p_C+p_R+p_H$. Thus, this is the time it takes for the dissipative processes to suppress coherent dynamics in the system (analogous to a T2-time for the machine in the language of spin relaxation; the time to equilibrate to the steady state would be a T1-time). If the sum of the bath couplings exceeds $g$, no oscillations are observed, as apparent from the red curves in \figref{fig.transplots}, while for large $g$ the imaginary part of $\lambda_{cp}$ is $2g$ as expected.

After coherent dynamics is damped out, each quantity approaches its steady-state value. The trace distance initially drops fast, in the coherent regime, and then follows an exponential decay. The rate of this approach to the steady state at long times is given by $\lambda_{max}$. Denoting the rewrapping of $\mathbf{v}$ to a density matrix $\rho(\mathbf{v})$ this can be seen easily
\begin{align}
D(\rho(t),\rho_\infty) & = \frac{1}{2} ||\rho(\mathbf{v}_h(t) + \mathbf{v}_\infty) - \rho(\mathbf{v}_\infty) ||_1 \nonumber \\
& = \frac{1}{2} ||\rho(\mathbf{v}_h(t))||_1 = \frac{1}{2} ||\rho(\sum_j c_j e^{\lambda_j t} \mathbf{e}_j)||_1 \nonumber \\
& \approx  \frac{1}{2} ||\rho(c_{max} e^{\lambda_{max} t} \mathbf{e}_{max})||_1 \nonumber \\
& = \frac{1}{2} c_{max} e^{\lambda_{max} t} ||\rho(\mathbf{e}_{max})||_1 ,
\end{align}
where $c_{max}$ and $\mathbf{e}_{max}$ are the coefficient and eigenvector corresponding to $\lambda_{max}$, and we used that the wrapping is linear and \eqref{eq.vht}. In general, $\lambda_{max}$ may depend on all the bath couplings, the interaction strength, and the temperatures. However, we observe that if the two smaller $p_i$ are equal, then $\lambda_{max}$ is equal to this value, $\lambda_{max} = \min\{p_C,p_R,p_H\}$, independent of $g$ and the temperatures. This is also the value of $\lambda_{max}$ if $g$ is small compared to the $p_i$. For large $g$, we have $\lambda_{max} \approx p_{min}+p_{min}'/4$, where $p_{min}$, $p_{min}'$ denote the two smallest couplings. Typical examples illustrating this behaviour are shown in \figref{fig.lambdag}(b). 

We note that it can be helpful to think of the machine as analogous to a damped oscillator, with the regimes where coherent evolution is visible or suppressed corresponding to under- and overdamping respectively.

\subsection{Cooling}

\begin{figure}[t]
\includegraphics[width=0.99\linewidth]{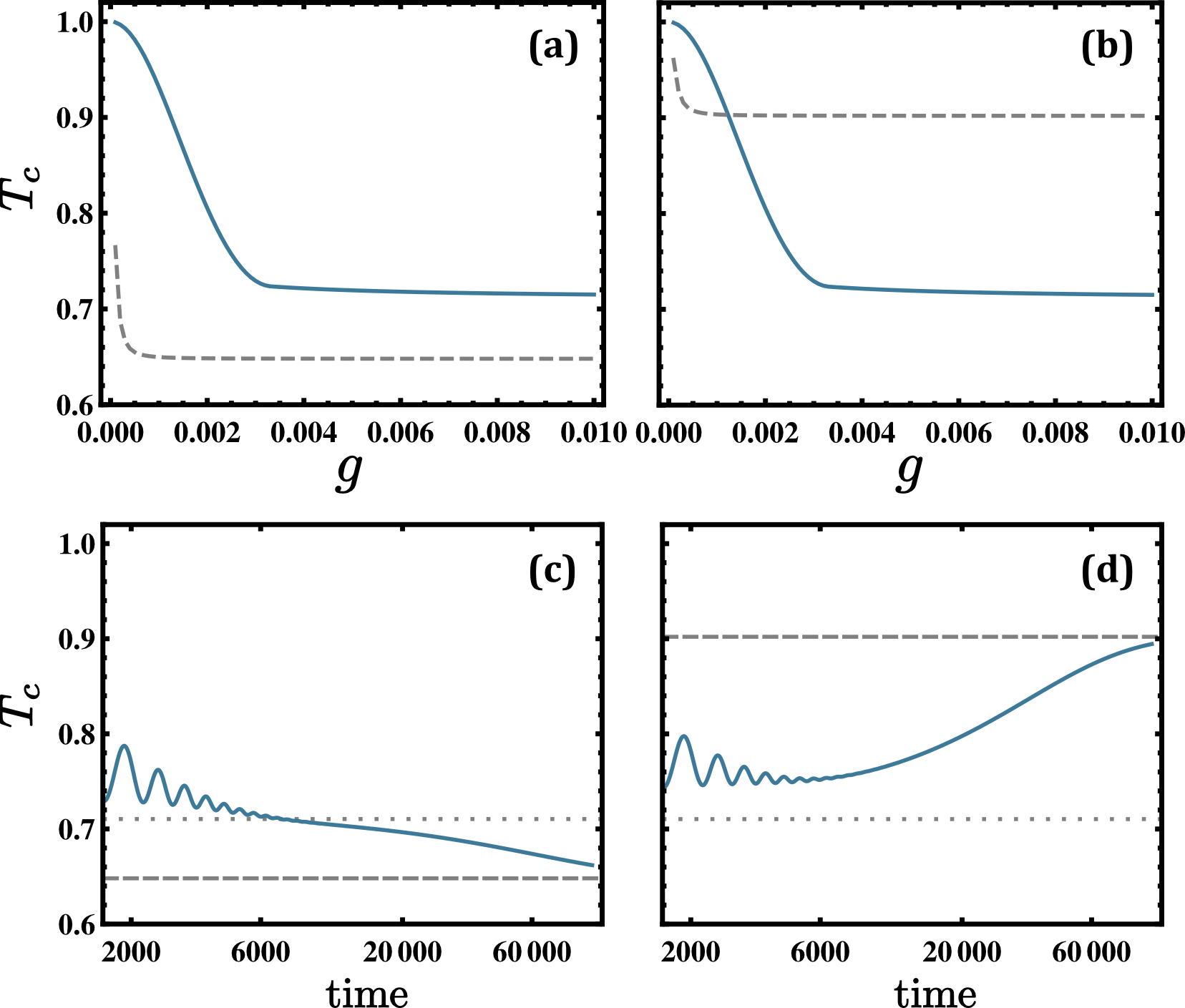}
\caption{(Color online) \textbf{(a)} Cold qubit temperature vs.~interaction strength. The lowest cold qubit temperature attained within a fixed time of $t=500$ is shown (solid) as well as the steady-state value (dashed), for $E_C=1$, $E_H=100$, $T_C=T_R=1$, $T_H=100$, $p_C=p_H=10^{-5}$, and $p_R=10^{-3}$. \textbf{(c)} Cold qubit temperature vs.~time corresponding to the bath couplings of (a) and $g=5\times 10^{-3}$. The lowest temperature attainable from the initial state by coherent dynamics is indicated (dotted) as well as the steady-state value (dashed). \textbf{(b)} and \textbf{(d)} Same as (a) and (c), with $p_C=10^{-4}$. Interestingly, the transient temperature of the cold qubit is significantly lower compared to the steady-state value.}
\label{fig.Tc_trans_inf}
\end{figure}

Both steady-state and transient cooling increases with increasing interaction strength. When the goal is to reach a low temperature for the cold qubit, it is always optimal to make $g$ as large as possible. The time it takes to cool, however, depends strongly on whether the system displays coherent dynamics or not, i.e.~on whether $g$ exceeds the $p_i$, and depending on the bath couplings, it can happen that colder temperatures are reached in the transient than in the steady-state regime, as was also noted in \cite{mitchison15}.

The difference between evolution with and without the coherent dynamics damped out is apparent from the examples in \figref{fig.transplots}. In \figref{fig.Tc_trans_inf}(a) we show the coldest temperature of the cold qubit reached within a fixed evolution time as a function of $g$. We see that as $g$ surpasses the bath couplings, there is a sharp drop in the cold qubit temperature attained at short times. The system cools faster when the interaction is strong such that the initial dynamics are coherent, as might be expected.

To separate out the effect of the dissipative evolution, we can compute the lowest cold qubit temperature which can be reached from the initial thermal state $\tau$ under unitary evolution alone with the Hamiltonian $H_0 + H_{int}$. The minimal temperature is reached when the ground state population of the cold qubit is maximal. This population (referring to the computational basis) is $\rho_{1,1}+\rho_{2,2}+\rho_{3,3}+\rho_{4,4}$, and since only $\rho_{3,3}$ is affected by the evolution, that means $\rho_{3,3}$ should be maximal. The evolution shuffles population between the states $|010\rangle$ and $|101\rangle$, corresponding to $\rho_{3,3}$ and $\rho_{6,6}$ and hence the maximal value of $\rho_{3,3}$ is $\max\,\{\tau_{3,3},\tau_{6,6}\}$. Which of these elements is larger depends on the bath temperatures and the energies. However, in a regime where the machine provides steady-state cooling, the larger one is always $\tau_{6,6}$. This can be seen using the virtual qubit picture \cite{brunner12}. Expressed in terms of the cold and (normalised) virtual qubit ground state populations at time zero, we have $\tau_{3,3} = r_C(1-r_V)$ and $\tau_{6,6} = (1-r_C) r_V$. Steady-state cooling requires $T_V < T_C$ \cite{brunner12} and hence $r_C < r_V$. Thus the minimal cold qubit temperature is reached when the populations of $|010\rangle$ and $|101\rangle$ are completely reversed, which happens after half a period at time $t=\pi/2g$.

Interestingly, due to the coherent dynamics, the cold qubit temperature in the transient regime can be significantly lower than the steady state value. As the initial dynamics is effectively dissipation free, the minimal temperature in this regime is roughly independent of the bath couplings, contrary to the steady-state value. An example of this behaviour is shown in \figref{fig.Tc_trans_inf}. In particular \figref{fig.Tc_trans_inf}(d) shows how a cold temperature is quickly reached during the coherent phase of the evolution, while after damping out of the oscillations, the temperature then approaches a higher steady-state value. In this case, optimal cooling is achieved at $t=\pi/2g$. 

\begin{figure}[t]
\includegraphics[width=0.80\linewidth]{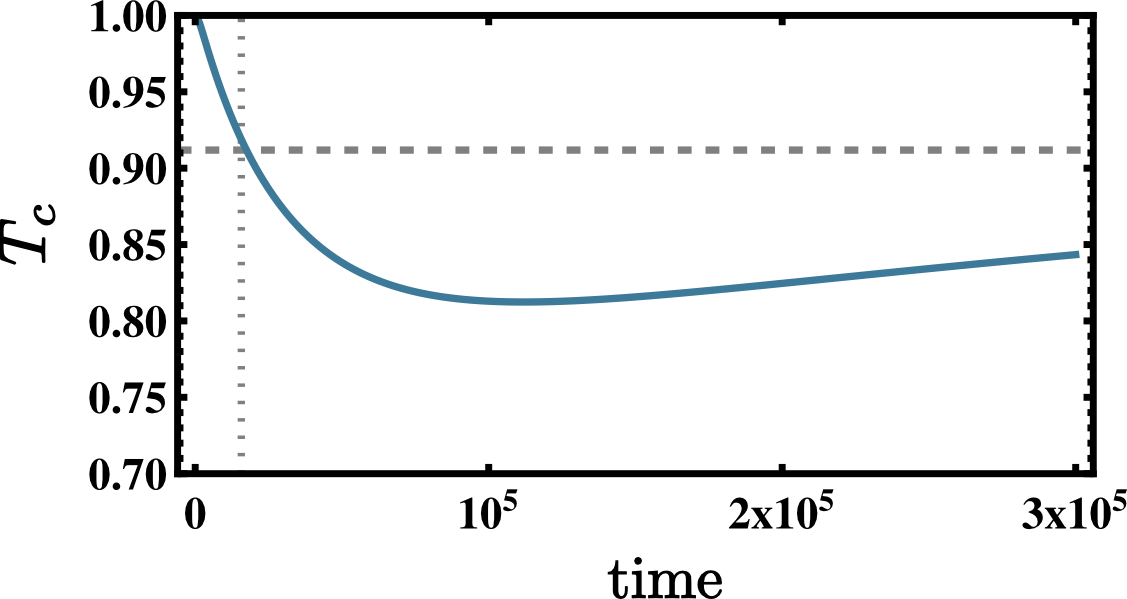}
\caption{(Color online) A minimum cold qubit temperature is reached in finite time for $E_C=1$, $E_H=100$, $T_C=T_R=1$, $T_H=100$, $p_C=10^{-5}$, $p_R=10^{-3}$, $p_H=10^{-5}$, $g=10^{-4}$. The steady-state temperature is indicated (dashed) and half a period $\pi/2g$ of coherent evolution without dissipation (dotted).}
\label{fig.Tc_min_fintime}
\end{figure}

While this optimal cooling would require a precise timing, we note however that even if control on this short time scale is not available, it may still be possible to achieve enhanced cooling by extracting the qubit before reaching the steady state. While coherent oscillations are damped out at a time set by the largest bath coupling, the approach rate to the steady state is determined by the smaller bath couplings, as discussed above. Hence when the couplings are different, there is an intermediate regime between coherent dynamics and steady state where the qubit temperature is low and precise time control is not necessary to extract it (c.f.~\figref{fig.Tc_trans_inf}(d)). A minimum cold qubit temperature can occur in finite time even when coherent dynamics is not visible. This can happen e.g.~when the coupling to the hot bath is very weak, while the the largest bath coupling exceeds the interaction strength, damping out oscillations. An example is shown in \figref{fig.Tc_min_fintime}. The optimal time in this case scales linearly with $p_R$ and inversely with $g$ but occurs much later than the half-period $\pi/2g$ corresponding to the first minimum under coherent dynamics. One can understand this, as well as the case where some initial oscillations are visible, in the picture of an underdamped oscillator approaching critical damping. As the damping is increased, in addition to the amplitude of oscillations decreasing their period increases and finally diverges as critical damping is reached.

We note that the effect of reaching a minimal temperature in finite time is robust to small deviations in the coupling and interaction parameters. The qualitative behaviour persists as long as the ordering of their magnitudes is preserved. Hence very precise control over the couplings is also not required to achieve enhanced cooling in the transient regime.

\subsection{Entanglement in the transient regime}

We can thus characterize the dynamics of entanglement. Here we focus on genuine tripartite entanglement, and on entanglement on the bipartition $R|CH$, i.e. the bipartition of energy in vs energy out. Examples are given in \figref{fig.transplots}(e)-(h). It is clearly seen that in the coherent regime, the amount of both types of entanglement can be considerably larger than in the steady state. 

In fact, it is simple to estimate how much entanglement can be created in the transient regime. This amount will depend on the relative sizes of the interaction strength and the bath couplings, as well as on the initial state. However if the interaction strength is larger than the bath couplings, at short times the evolution is approximately dissipation free, as mentioned above. The system thus evolves approximately unitarily under the Hamiltonian $H_0 + H_{int}$, and the maximal entanglement is what can be generated from the initial state $\tau$ under this evolution. The only elements of $\tau$ which can change are those in the degenerate subspace affected by $H_{int}$. The second term in \eqref{eq.entwitness} is then invariant and the maximal entanglement is obtained by maximising $|\rho_{3,6}|$ which occurs after a quarter period at $t = \pi/4g$. The optimal value is $|\rho_{3,6}| = |\tau_{3,3}-\tau_{6,6}|$ which results in a maximal bipartite entanglement of 
\begin{align}
\label{eq.maxW}
W_{R| CH}^{max} & =  N^{-1} |e^\frac{E_C+E_H}{T_R} - e^{\frac{E_C}{T_C}+\frac{E_H}{T_H}}| \nonumber \\
& - 2 N^{-1} e^{\frac{E_C}{2}\left(\frac{1}{T_C}+\frac{1}{T_R}\right) + \frac{E_H}{2}\left(\frac{1}{T_R}+\frac{1}{T_H}\right)} ,
\end{align}
where $N = (1+e^{E_C/T_C})(1+e^{E_R/T_R})(1+e^{E_H/T_H})$. The maximum is the same across the other bipartitions, however we note that once the thermal dissipation is accounted for, entanglement on $R|CH$ appears to dominate the other two. For genuine tripartite entanglement, $W_{R CH}^{max}$ is obtained from \eqref{eq.maxW} by replacing replacing the factor 2 with 6 in the negative term.

As an example, the maximal bipartite and genuine tripartite entanglement extractable by the unitary dynamics is indicated on \figref{fig.transplots}(e)-(h). It represents an upper bound on the entanglement which can be extracted when one has control over the system at the fast time scale set by $g$. As can be seen, the maximal value is never reached exactly, due to the presence of the bath couplings. While entanglement in the transient regime is maximised for interaction strength large compared to the bath couplings, this is not optimal for generation of steady-state entanglement. Unlike for cooling, it is not optimal to make $g$ as large as possible. As a result, when optimising for steady-state entanglement, no oscillations are observed in the transient regime, and the entanglement remains small compared with the largest value obtainable from dissipation free evolution.

\section{Conclusion}

The dynamics of a small quantum absorption refrigerator was investigated. We discussed the approach to equilibrium, in particular with respect to time scales, cooling, and entanglement. Notably, we observed that optimal cooling may occur at finite times, and not in the steady-state regime. Similarly, the largest amount of entanglement is usually obtained in the transient regime. As pointed recently in Ref. \cite{mitchison15}, this highlights the relevance of quantum effects in such thermal machines.

We believe that the observation that enhanced cooling can be achieved in the transient regime without the need for precise timing or control of the fridge parameters may open interesting possibilities for cooling or initializing quantum systems. 

More generally, the methods discussed in the present work could be adapted to other models of quantum refrigerators, as well as to other quantum thermal machines, producing work or entanglement \cite{brask15}.

\emph{Acknowledgements.} We thank G\'eraldine Haack, Marcus Huber, Ralph Silva and Paul Skrzypczyk for discussions. We acknowledge financial support from the Swiss National Science Foundation (grant PP00P2\_138917, QSIT, and Starting Grant DIAQ), and SEFRI (COST action MP1006).

\emph{Note added.} While finishing this work, we became aware of a related work by Mitchison and colleagues \cite{mitchison15}.


\begin{thebibliography}{28}%
\makeatletter
\providecommand \@ifxundefined [1]{%
 \@ifx{#1\undefined}
}%
\providecommand \@ifnum [1]{%
 \ifnum #1\expandafter \@firstoftwo
 \else \expandafter \@secondoftwo
 \fi
}%
\providecommand \@ifx [1]{%
 \ifx #1\expandafter \@firstoftwo
 \else \expandafter \@secondoftwo
 \fi
}%
\providecommand \natexlab [1]{#1}%
\providecommand \enquote  [1]{``#1''}%
\providecommand \bibnamefont  [1]{#1}%
\providecommand \bibfnamefont [1]{#1}%
\providecommand \citenamefont [1]{#1}%
\providecommand \href@noop [0]{\@secondoftwo}%
\providecommand \href [0]{\begingroup \@sanitize@url \@href}%
\providecommand \@href[1]{\@@startlink{#1}\@@href}%
\providecommand \@@href[1]{\endgroup#1\@@endlink}%
\providecommand \@sanitize@url [0]{\catcode `\\12\catcode `\$12\catcode
  `\&12\catcode `\#12\catcode `\^12\catcode `\_12\catcode `\%12\relax}%
\providecommand \@@startlink[1]{}%
\providecommand \@@endlink[0]{}%
\providecommand \url  [0]{\begingroup\@sanitize@url \@url }%
\providecommand \@url [1]{\endgroup\@href {#1}{\urlprefix }}%
\providecommand \urlprefix  [0]{URL }%
\providecommand \Eprint [0]{\href }%
\providecommand \doibase [0]{http://dx.doi.org/}%
\providecommand \selectlanguage [0]{\@gobble}%
\providecommand \bibinfo  [0]{\@secondoftwo}%
\providecommand \bibfield  [0]{\@secondoftwo}%
\providecommand \translation [1]{[#1]}%
\providecommand \BibitemOpen [0]{}%
\providecommand \bibitemStop [0]{}%
\providecommand \bibitemNoStop [0]{.\EOS\space}%
\providecommand \EOS [0]{\spacefactor3000\relax}%
\providecommand \BibitemShut  [1]{\csname bibitem#1\endcsname}%
\let\auto@bib@innerbib\@empty
\bibitem [{\citenamefont {Kosloff}\ and\ \citenamefont {Levy}(2014)}]{review1}%
  \BibitemOpen
  \bibfield  {author} {\bibinfo {author} {\bibfnamefont {R.}~\bibnamefont
  {Kosloff}}\ and\ \bibinfo {author} {\bibfnamefont {A.}~\bibnamefont {Levy}},\
  }\href {\doibase 10.1146/annurev-physchem-040513-103724} {\bibfield
  {journal} {\bibinfo  {journal} {Annual Review of Physical Chemistry}\
  }\textbf {\bibinfo {volume} {65}},\ \bibinfo {pages} {365} (\bibinfo {year}
  {2014})},\ \bibinfo {note} {pMID: 24689798},\ \Eprint
  {http://arxiv.org/abs/http://dx.doi.org/10.1146/annurev-physchem-040513-103724}
  {http://dx.doi.org/10.1146/annurev-physchem-040513-103724} \BibitemShut
  {NoStop}%
\bibitem [{\citenamefont {Gelbwaser-Klimovsky}\ \emph
  {et~al.}(2015)\citenamefont {Gelbwaser-Klimovsky}, \citenamefont {Niedenzu},\
  and\ \citenamefont {Kurizki}}]{review2}%
  \BibitemOpen
  \bibfield  {author} {\bibinfo {author} {\bibfnamefont {D.}~\bibnamefont
  {Gelbwaser-Klimovsky}}, \bibinfo {author} {\bibfnamefont {W.}~\bibnamefont
  {Niedenzu}}, \ and\ \bibinfo {author} {\bibfnamefont {G.}~\bibnamefont
  {Kurizki}},\ }\href {http://arxiv.org/abs/1503.01195} {\bibfield  {journal}
  {\bibinfo  {journal} {arXiv e-print}\ ,\ \bibinfo {pages} {1503.01195}}
  (\bibinfo {year} {2015})}\BibitemShut {NoStop}%
\bibitem [{\citenamefont {Goold}\ \emph {et~al.}(2015)\citenamefont {Goold},
  \citenamefont {Huber}, \citenamefont {Riera}, \citenamefont {del Rio},\ and\
  \citenamefont {Skrzypczyk}}]{Goold15}%
  \BibitemOpen
  \bibfield  {author} {\bibinfo {author} {\bibfnamefont {J.}~\bibnamefont
  {Goold}}, \bibinfo {author} {\bibfnamefont {M.}~\bibnamefont {Huber}},
  \bibinfo {author} {\bibfnamefont {A.}~\bibnamefont {Riera}}, \bibinfo
  {author} {\bibfnamefont {L.}~\bibnamefont {del Rio}}, \ and\ \bibinfo
  {author} {\bibfnamefont {P.}~\bibnamefont {Skrzypczyk}},\ }\href
  {http://arxiv.org/abs/1505.07835} {\bibfield  {journal} {\bibinfo  {journal}
  {arXiv e-print}\ ,\ \bibinfo {pages} {1505.07835}} (\bibinfo {year}
  {2015})}\BibitemShut {NoStop}%
\bibitem [{\citenamefont {Vinjanampathy}\ and\ \citenamefont
  {Anders}(2015)}]{anders15}%
  \BibitemOpen
  \bibfield  {author} {\bibinfo {author} {\bibfnamefont {S.}~\bibnamefont
  {Vinjanampathy}}\ and\ \bibinfo {author} {\bibfnamefont {J.}~\bibnamefont
  {Anders}},\ }\href {http://arxiv.org/abs/1508.06099} {\bibfield  {journal}
  {\bibinfo  {journal} {arXiv [quant-ph] e-print}\ ,\ \bibinfo {pages}
  {1508.06099}} (\bibinfo {year} {2015})}\BibitemShut {NoStop}%
\bibitem [{\citenamefont {Scovil}\ and\ \citenamefont
  {Schulz-DuBois}(1959)}]{scovil59}%
  \BibitemOpen
  \bibfield  {author} {\bibinfo {author} {\bibfnamefont {H.~E.~D.}\
  \bibnamefont {Scovil}}\ and\ \bibinfo {author} {\bibfnamefont {E.~O.}\
  \bibnamefont {Schulz-DuBois}},\ }\href {\doibase 10.1103/PhysRevLett.2.262}
  {\bibfield  {journal} {\bibinfo  {journal} {Phys. Rev. Lett.}\ }\textbf
  {\bibinfo {volume} {2}},\ \bibinfo {pages} {262} (\bibinfo {year}
  {1959})}\BibitemShut {NoStop}%
\bibitem [{\citenamefont {Geva}\ and\ \citenamefont
  {Kosloff}(1996)}]{GevKos96}%
  \BibitemOpen
  \bibfield  {author} {\bibinfo {author} {\bibfnamefont {E.}~\bibnamefont
  {Geva}}\ and\ \bibinfo {author} {\bibfnamefont {R.}~\bibnamefont {Kosloff}},\
  }\href {\doibase http://dx.doi.org/10.1063/1.471453} {\bibfield  {journal}
  {\bibinfo  {journal} {The Journal of Chemical Physics}\ }\textbf {\bibinfo
  {volume} {104}},\ \bibinfo {pages} {7681} (\bibinfo {year}
  {1996})}\BibitemShut {NoStop}%
\bibitem [{\citenamefont {Palao}\ \emph {et~al.}(2001)\citenamefont {Palao},
  \citenamefont {Kosloff},\ and\ \citenamefont {Gordon}}]{PalKosGor01}%
  \BibitemOpen
  \bibfield  {author} {\bibinfo {author} {\bibfnamefont {J.~P.}\ \bibnamefont
  {Palao}}, \bibinfo {author} {\bibfnamefont {R.}~\bibnamefont {Kosloff}}, \
  and\ \bibinfo {author} {\bibfnamefont {J.~M.}\ \bibnamefont {Gordon}},\
  }\href {\doibase 10.1103/PhysRevE.64.056130} {\bibfield  {journal} {\bibinfo
  {journal} {Phys. Rev. E}\ }\textbf {\bibinfo {volume} {64}},\ \bibinfo
  {pages} {056130} (\bibinfo {year} {2001})}\BibitemShut {NoStop}%
\bibitem [{\citenamefont {Correa}(2014)}]{luis}%
  \BibitemOpen
  \bibfield  {author} {\bibinfo {author} {\bibfnamefont {L.~A.}\ \bibnamefont
  {Correa}},\ }\href {\doibase 10.1103/PhysRevE.89.042128} {\bibfield
  {journal} {\bibinfo  {journal} {Phys. Rev. E}\ }\textbf {\bibinfo {volume}
  {89}},\ \bibinfo {pages} {042128} (\bibinfo {year} {2014})}\BibitemShut
  {NoStop}%
\bibitem [{\citenamefont {Silva}\ \emph {et~al.}(2015)\citenamefont {Silva},
  \citenamefont {Skrzypczyk},\ and\ \citenamefont {Brunner}}]{ralph}%
  \BibitemOpen
  \bibfield  {author} {\bibinfo {author} {\bibfnamefont {R.}~\bibnamefont
  {Silva}}, \bibinfo {author} {\bibfnamefont {P.}~\bibnamefont {Skrzypczyk}}, \
  and\ \bibinfo {author} {\bibfnamefont {N.}~\bibnamefont {Brunner}},\ }\href
  {\doibase 10.1103/PhysRevE.92.012136} {\bibfield  {journal} {\bibinfo
  {journal} {Phys. Rev. E}\ }\textbf {\bibinfo {volume} {92}},\ \bibinfo
  {pages} {012136} (\bibinfo {year} {2015})}\BibitemShut {NoStop}%
\bibitem [{\citenamefont {Linden}\ \emph {et~al.}(2010)\citenamefont {Linden},
  \citenamefont {Popescu},\ and\ \citenamefont {Skrzypczyk}}]{linden10}%
  \BibitemOpen
  \bibfield  {author} {\bibinfo {author} {\bibfnamefont {N.}~\bibnamefont
  {Linden}}, \bibinfo {author} {\bibfnamefont {S.}~\bibnamefont {Popescu}}, \
  and\ \bibinfo {author} {\bibfnamefont {P.}~\bibnamefont {Skrzypczyk}},\
  }\href {\doibase 10.1103/PhysRevLett.105.130401} {\bibfield  {journal}
  {\bibinfo  {journal} {Phys. Rev. Lett.}\ }\textbf {\bibinfo {volume} {105}},\
  \bibinfo {pages} {130401} (\bibinfo {year} {2010})}\BibitemShut {NoStop}%
\bibitem [{\citenamefont {Skrzypczyk}\ \emph {et~al.}(2011)\citenamefont
  {Skrzypczyk}, \citenamefont {Brunner}, \citenamefont {Linden},\ and\
  \citenamefont {Popescu}}]{skrzypczyk11}%
  \BibitemOpen
  \bibfield  {author} {\bibinfo {author} {\bibfnamefont {P.}~\bibnamefont
  {Skrzypczyk}}, \bibinfo {author} {\bibfnamefont {N.}~\bibnamefont {Brunner}},
  \bibinfo {author} {\bibfnamefont {N.}~\bibnamefont {Linden}}, \ and\ \bibinfo
  {author} {\bibfnamefont {S.}~\bibnamefont {Popescu}},\ }\href
  {http://stacks.iop.org/1751-8121/44/i=49/a=492002} {\bibfield  {journal}
  {\bibinfo  {journal} {Journal of Physics A: Mathematical and Theoretical}\
  }\textbf {\bibinfo {volume} {44}},\ \bibinfo {pages} {492002} (\bibinfo
  {year} {2011})}\BibitemShut {NoStop}%
\bibitem [{\citenamefont {Correa}\ \emph {et~al.}(2013)\citenamefont {Correa},
  \citenamefont {Palao}, \citenamefont {Adesso},\ and\ \citenamefont
  {Alonso}}]{correa13}%
  \BibitemOpen
  \bibfield  {author} {\bibinfo {author} {\bibfnamefont {L.~A.}\ \bibnamefont
  {Correa}}, \bibinfo {author} {\bibfnamefont {J.~P.}\ \bibnamefont {Palao}},
  \bibinfo {author} {\bibfnamefont {G.}~\bibnamefont {Adesso}}, \ and\ \bibinfo
  {author} {\bibfnamefont {D.}~\bibnamefont {Alonso}},\ }\href {\doibase
  10.1103/PhysRevE.87.042131} {\bibfield  {journal} {\bibinfo  {journal} {Phys.
  Rev. E}\ }\textbf {\bibinfo {volume} {87}},\ \bibinfo {pages} {042131}
  (\bibinfo {year} {2013})}\BibitemShut {NoStop}%
\bibitem [{\citenamefont {Woods}\ \emph {et~al.}(2015)\citenamefont {Woods},
  \citenamefont {Ng},\ and\ \citenamefont {Wehner}}]{woods15}%
  \BibitemOpen
  \bibfield  {author} {\bibinfo {author} {\bibfnamefont {M.~P.}\ \bibnamefont
  {Woods}}, \bibinfo {author} {\bibfnamefont {N.}~\bibnamefont {Ng}}, \ and\
  \bibinfo {author} {\bibfnamefont {S.}~\bibnamefont {Wehner}},\ }\href
  {http://arxiv.org/abs/1506.02322} {\bibfield  {journal} {\bibinfo  {journal}
  {arXiv [quant-ph] e-print}\ ,\ \bibinfo {pages} {1506.02322}} (\bibinfo
  {year} {2015})}\BibitemShut {NoStop}%
\bibitem [{\citenamefont {Brunner}\ \emph {et~al.}(2012)\citenamefont
  {Brunner}, \citenamefont {Linden}, \citenamefont {Popescu},\ and\
  \citenamefont {Skrzypczyk}}]{brunner12}%
  \BibitemOpen
  \bibfield  {author} {\bibinfo {author} {\bibfnamefont {N.}~\bibnamefont
  {Brunner}}, \bibinfo {author} {\bibfnamefont {N.}~\bibnamefont {Linden}},
  \bibinfo {author} {\bibfnamefont {S.}~\bibnamefont {Popescu}}, \ and\
  \bibinfo {author} {\bibfnamefont {P.}~\bibnamefont {Skrzypczyk}},\ }\href
  {\doibase 10.1103/PhysRevE.85.051117} {\bibfield  {journal} {\bibinfo
  {journal} {Phys. Rev. E}\ }\textbf {\bibinfo {volume} {85}},\ \bibinfo
  {pages} {051117} (\bibinfo {year} {2012})}\BibitemShut {NoStop}%
\bibitem [{\citenamefont {Brunner}\ \emph {et~al.}(2014)\citenamefont
  {Brunner}, \citenamefont {Huber}, \citenamefont {Linden}, \citenamefont
  {Popescu}, \citenamefont {Silva},\ and\ \citenamefont
  {Skrzypczyk}}]{brunner14}%
  \BibitemOpen
  \bibfield  {author} {\bibinfo {author} {\bibfnamefont {N.}~\bibnamefont
  {Brunner}}, \bibinfo {author} {\bibfnamefont {M.}~\bibnamefont {Huber}},
  \bibinfo {author} {\bibfnamefont {N.}~\bibnamefont {Linden}}, \bibinfo
  {author} {\bibfnamefont {S.}~\bibnamefont {Popescu}}, \bibinfo {author}
  {\bibfnamefont {R.}~\bibnamefont {Silva}}, \ and\ \bibinfo {author}
  {\bibfnamefont {P.}~\bibnamefont {Skrzypczyk}},\ }\href {\doibase
  10.1103/PhysRevE.89.032115} {\bibfield  {journal} {\bibinfo  {journal} {Phys.
  Rev. E}\ }\textbf {\bibinfo {volume} {89}},\ \bibinfo {pages} {032115}
  (\bibinfo {year} {2014})}\BibitemShut {NoStop}%
\bibitem [{\citenamefont {Chen}\ and\ \citenamefont {Li}(2012)}]{chen12}%
  \BibitemOpen
  \bibfield  {author} {\bibinfo {author} {\bibfnamefont {Y.-X.}\ \bibnamefont
  {Chen}}\ and\ \bibinfo {author} {\bibfnamefont {S.-W.}\ \bibnamefont {Li}},\
  }\href {http://stacks.iop.org/0295-5075/97/i=4/a=40003} {\bibfield  {journal}
  {\bibinfo  {journal} {EPL (Europhysics Letters)}\ }\textbf {\bibinfo {volume}
  {97}},\ \bibinfo {pages} {40003} (\bibinfo {year} {2012})}\BibitemShut
  {NoStop}%
\bibitem [{\citenamefont {Venturelli}\ \emph {et~al.}(2013)\citenamefont
  {Venturelli}, \citenamefont {Fazio},\ and\ \citenamefont
  {Giovannetti}}]{venturelli13}%
  \BibitemOpen
  \bibfield  {author} {\bibinfo {author} {\bibfnamefont {D.}~\bibnamefont
  {Venturelli}}, \bibinfo {author} {\bibfnamefont {R.}~\bibnamefont {Fazio}}, \
  and\ \bibinfo {author} {\bibfnamefont {V.}~\bibnamefont {Giovannetti}},\
  }\href {\doibase 10.1103/PhysRevLett.110.256801} {\bibfield  {journal}
  {\bibinfo  {journal} {Phys. Rev. Lett.}\ }\textbf {\bibinfo {volume} {110}},\
  \bibinfo {pages} {256801} (\bibinfo {year} {2013})}\BibitemShut {NoStop}%
\bibitem [{\citenamefont {Bellomo}\ \emph {et~al.}(2008)\citenamefont
  {Bellomo}, \citenamefont {Lo~Franco}, \citenamefont {Maniscalco},\ and\
  \citenamefont {Compagno}}]{bellomo}%
  \BibitemOpen
  \bibfield  {author} {\bibinfo {author} {\bibfnamefont {B.}~\bibnamefont
  {Bellomo}}, \bibinfo {author} {\bibfnamefont {R.}~\bibnamefont {Lo~Franco}},
  \bibinfo {author} {\bibfnamefont {S.}~\bibnamefont {Maniscalco}}, \ and\
  \bibinfo {author} {\bibfnamefont {G.}~\bibnamefont {Compagno}},\ }\href
  {\doibase 10.1103/PhysRevA.78.060302} {\bibfield  {journal} {\bibinfo
  {journal} {Phys. Rev. A}\ }\textbf {\bibinfo {volume} {78}},\ \bibinfo
  {pages} {060302} (\bibinfo {year} {2008})}\BibitemShut {NoStop}%
\bibitem [{\citenamefont {Mitchison}\ \emph {et~al.}(2015)\citenamefont
  {Mitchison}, \citenamefont {Woods}, \citenamefont {Prior},\ and\
  \citenamefont {Huber}}]{mitchison15}%
  \BibitemOpen
  \bibfield  {author} {\bibinfo {author} {\bibfnamefont {M.~T.}\ \bibnamefont
  {Mitchison}}, \bibinfo {author} {\bibfnamefont {M.~P.}\ \bibnamefont
  {Woods}}, \bibinfo {author} {\bibfnamefont {J.}~\bibnamefont {Prior}}, \ and\
  \bibinfo {author} {\bibfnamefont {M.}~\bibnamefont {Huber}},\ }\href
  {http://arxiv.org/abs/1005.2114} {\bibfield  {journal} {\bibinfo  {journal}
  {arXiv [quant-ph] e-print}\ ,\ \bibinfo {pages} {1504.01593}} (\bibinfo
  {year} {2015})}\BibitemShut {NoStop}%
\bibitem [{\citenamefont {Palma}\ \emph {et~al.}(1996)\citenamefont {Palma},
  \citenamefont {Suominen},\ and\ \citenamefont {Ekert}}]{palma96}%
  \BibitemOpen
  \bibfield  {author} {\bibinfo {author} {\bibfnamefont {G.~M.}\ \bibnamefont
  {Palma}}, \bibinfo {author} {\bibfnamefont {K.-A.}\ \bibnamefont {Suominen}},
  \ and\ \bibinfo {author} {\bibfnamefont {A.~K.}\ \bibnamefont {Ekert}},\
  }\href {\doibase 10.1098/rspa.1996.0029} {\bibfield  {journal} {\bibinfo
  {journal} {Proceedings of the Royal Society of London A: Mathematical,
  Physical and Engineering Sciences}\ }\textbf {\bibinfo {volume} {452}},\
  \bibinfo {pages} {567} (\bibinfo {year} {1996})}\BibitemShut {NoStop}%
\bibitem [{\citenamefont {Nielsen}\ and\ \citenamefont
  {Chuang}(2007)}]{nielsenchuang}%
  \BibitemOpen
  \bibfield  {author} {\bibinfo {author} {\bibfnamefont {M.~A.}\ \bibnamefont
  {Nielsen}}\ and\ \bibinfo {author} {\bibfnamefont {I.~L.}\ \bibnamefont
  {Chuang}},\ }\href@noop {} {\emph {\bibinfo {title} {Quantum Computation and
  Quantum Information}}}\ (\bibinfo  {publisher} {Cambridge University Press},\
  \bibinfo {address} {Cambridge, UK},\ \bibinfo {year} {2007})\BibitemShut
  {NoStop}%
\bibitem [{\citenamefont {Gühne}\ and\ \citenamefont
  {Seevinck}(2010)}]{guhne10}%
  \BibitemOpen
  \bibfield  {author} {\bibinfo {author} {\bibfnamefont {O.}~\bibnamefont
  {Gühne}}\ and\ \bibinfo {author} {\bibfnamefont {M.}~\bibnamefont
  {Seevinck}},\ }\href {http://stacks.iop.org/1367-2630/12/i=5/a=053002}
  {\bibfield  {journal} {\bibinfo  {journal} {New Journal of Physics}\ }\textbf
  {\bibinfo {volume} {12}},\ \bibinfo {pages} {053002} (\bibinfo {year}
  {2010})}\BibitemShut {NoStop}%
\bibitem [{\citenamefont {Huber}\ \emph {et~al.}(2010)\citenamefont {Huber},
  \citenamefont {Mintert}, \citenamefont {Gabriel},\ and\ \citenamefont
  {Hiesmayr}}]{huber10}%
  \BibitemOpen
  \bibfield  {author} {\bibinfo {author} {\bibfnamefont {M.}~\bibnamefont
  {Huber}}, \bibinfo {author} {\bibfnamefont {F.}~\bibnamefont {Mintert}},
  \bibinfo {author} {\bibfnamefont {A.}~\bibnamefont {Gabriel}}, \ and\
  \bibinfo {author} {\bibfnamefont {B.~C.}\ \bibnamefont {Hiesmayr}},\ }\href
  {\doibase 10.1103/PhysRevLett.104.210501} {\bibfield  {journal} {\bibinfo
  {journal} {Phys. Rev. Lett.}\ }\textbf {\bibinfo {volume} {104}},\ \bibinfo
  {pages} {210501} (\bibinfo {year} {2010})}\BibitemShut {NoStop}%
\bibitem [{\citenamefont {Wootters}(1998)}]{wootters}%
  \BibitemOpen
  \bibfield  {author} {\bibinfo {author} {\bibfnamefont {W.~K.}\ \bibnamefont
  {Wootters}},\ }\href {\doibase 10.1103/PhysRevLett.80.2245} {\bibfield
  {journal} {\bibinfo  {journal} {Phys. Rev. Lett.}\ }\textbf {\bibinfo
  {volume} {80}},\ \bibinfo {pages} {2245} (\bibinfo {year}
  {1998})}\BibitemShut {NoStop}%
\bibitem [{\citenamefont {Ma}\ \emph {et~al.}(2011)\citenamefont {Ma},
  \citenamefont {Chen}, \citenamefont {Chen}, \citenamefont {Spengler},
  \citenamefont {Gabriel},\ and\ \citenamefont {Huber}}]{ma11}%
  \BibitemOpen
  \bibfield  {author} {\bibinfo {author} {\bibfnamefont {Z.-H.}\ \bibnamefont
  {Ma}}, \bibinfo {author} {\bibfnamefont {Z.-H.}\ \bibnamefont {Chen}},
  \bibinfo {author} {\bibfnamefont {J.-L.}\ \bibnamefont {Chen}}, \bibinfo
  {author} {\bibfnamefont {C.}~\bibnamefont {Spengler}}, \bibinfo {author}
  {\bibfnamefont {A.}~\bibnamefont {Gabriel}}, \ and\ \bibinfo {author}
  {\bibfnamefont {M.}~\bibnamefont {Huber}},\ }\href {\doibase
  10.1103/PhysRevA.83.062325} {\bibfield  {journal} {\bibinfo  {journal} {Phys.
  Rev. A}\ }\textbf {\bibinfo {volume} {83}},\ \bibinfo {pages} {062325}
  (\bibinfo {year} {2011})}\BibitemShut {NoStop}%
\bibitem [{\citenamefont {Wu}\ \emph {et~al.}(2012)\citenamefont {Wu},
  \citenamefont {Kampermann}, \citenamefont {Bru\ss{}}, \citenamefont
  {Kl\"ockl},\ and\ \citenamefont {Huber}}]{wu12}%
  \BibitemOpen
  \bibfield  {author} {\bibinfo {author} {\bibfnamefont {J.-Y.}\ \bibnamefont
  {Wu}}, \bibinfo {author} {\bibfnamefont {H.}~\bibnamefont {Kampermann}},
  \bibinfo {author} {\bibfnamefont {D.}~\bibnamefont {Bru\ss{}}}, \bibinfo
  {author} {\bibfnamefont {C.}~\bibnamefont {Kl\"ockl}}, \ and\ \bibinfo
  {author} {\bibfnamefont {M.}~\bibnamefont {Huber}},\ }\href {\doibase
  10.1103/PhysRevA.86.022319} {\bibfield  {journal} {\bibinfo  {journal} {Phys.
  Rev. A}\ }\textbf {\bibinfo {volume} {86}},\ \bibinfo {pages} {022319}
  (\bibinfo {year} {2012})}\BibitemShut {NoStop}%
\bibitem [{\citenamefont {Hashemi~Rafsanjani}\ \emph
  {et~al.}(2012)\citenamefont {Hashemi~Rafsanjani}, \citenamefont {Huber},
  \citenamefont {Broadbent},\ and\ \citenamefont {Eberly}}]{hashemi12}%
  \BibitemOpen
  \bibfield  {author} {\bibinfo {author} {\bibfnamefont {S.~M.}\ \bibnamefont
  {Hashemi~Rafsanjani}}, \bibinfo {author} {\bibfnamefont {M.}~\bibnamefont
  {Huber}}, \bibinfo {author} {\bibfnamefont {C.~J.}\ \bibnamefont
  {Broadbent}}, \ and\ \bibinfo {author} {\bibfnamefont {J.~H.}\ \bibnamefont
  {Eberly}},\ }\href {\doibase 10.1103/PhysRevA.86.062303} {\bibfield
  {journal} {\bibinfo  {journal} {Phys. Rev. A}\ }\textbf {\bibinfo {volume}
  {86}},\ \bibinfo {pages} {062303} (\bibinfo {year} {2012})}\BibitemShut
  {NoStop}%
\bibitem [{\citenamefont {Brask}\ \emph {et~al.}(2015)\citenamefont {Brask},
  \citenamefont {Brunner}, \citenamefont {Haack},\ and\ \citenamefont
  {Huber}}]{brask15}%
  \BibitemOpen
  \bibfield  {author} {\bibinfo {author} {\bibfnamefont {J.~B.}\ \bibnamefont
  {Brask}}, \bibinfo {author} {\bibfnamefont {G.}~\bibnamefont {Haack}}, 
  \bibinfo {author} {\bibfnamefont {N.}~\bibnamefont {Brunner}}, \ and\ \bibinfo
  {author} {\bibfnamefont {M.}~\bibnamefont {Huber}},\ }\href
  {\doibase 10.1103/PhysRevE.92.062101} {\bibfield  {journal} {\bibinfo  {journal}
  {New J. Phys.}\ }\textbf {\bibinfo {volume}
  {17}},\ \bibinfo {pages} {113029} (\bibinfo {year}
  {2015})}\BibitemShut {NoStop}%
\end{thebibliography}

%

\end{document}